\documentclass[published,notoc]{JHEP} 

\JHEP{ sample file}

\usepackage{epsfig,multicol}

\usepackage{latexsym,amssymb}

\def\tilde{\widetilde}

\def\1bar{1\hskip -.275cm -}
\def\2bar{2\hskip -.275cm -}
\def\3bar{3\hskip -.275cm -}

\newsavebox{\uuunit}
\sbox{\uuunit}
    {\setlength{\unitlength}{0.825em}
     \begin{picture}(0.6,0.7)
        \thinlines
        \put(0,0){\line(1,0){0.5}}
        \put(0.15,0){\line(0,1){0.7}}
        \put(0.35,0){\line(0,1){0.8}}
       \multiput(0.3,0.8)(-0.04,-0.02){10}{\rule{0.5pt}{0.5pt}}
     \end {picture}}

\makeatletter \@addtoreset{equation}{section} \makeatother

\def\bfone{\relax{\rm 1\kern-.35em 1}}

\def\bfone{\relax{\rm 1\kern-.35em 1}}
 
\newcommand{\be}{\begin{equation}}
\newcommand{\ee}{\end{equation}}
\newcommand{\bea}{\begin{eqnarray}}
\newcommand{\eea}{\end{eqnarray}}

\newcommand{\nn}{\nonumber}

\title{ Walls from fluxes: An analytic RG-flow}

\author{Jose F. Morales and M. Trigiante\\
    Spinoza Institute, Postbus 80.195, 3508 TD Utrecht, The Netherlands.
    E-mail: \email{morales@phys.uu.nl}, \email{m.trigiante@phys.uu.nl}}

\preprint{\hepth{0112059}}  

\abstract{We construct supergravity solutions describing the near
horizon limit of D1D5 systems with non-trivial
 boundary conditions. Upon reduction to
five dimensions they define Melvin universes with
 NS--NS/RR fluxes, that smoothly interpolate
between two different $AdS_3$ geometries which define fixed points
for the RG--flow of the dual field theory.
We discuss the decoupling limits at the two ends of the flow.
 We also present a systematic study of the global properties of our
solution. In particular we show how, although the
$AdS_3\times S^3$ global isometry
group is broken down to $SU(2)_R\times U(1)^3$ by  global
identifications, a full two-dimensional conformal group
of isometries, with the expected central charge,
is restored at infinity.}


\begin{document}
%

\section{Introduction}
\label{intro}

The proposals \cite{maldacena,altri} for an holographic correspondence between
String (or M-) theory on Anti-de Sitter backgrounds
and suitable boundary gauge theories
have provided a powerful tool for the study of the strong
coupling regime of SYM theories.
Amazingly enough, phenomena like
confinement, chiral symmetry breaking and strong-weak
Seiberg dualities  in certain ${\cal N}=1$
gauge theories
are already described in terms of the dual classical supergravity \cite{ps}.
 These extremely encouraging results are
however limited to a very restricted number of examples.
A similar analysis in a
non--conformal ${\cal N}=2$ framework has been hampered so far
by the presence of
{\it enhan\c{c}on singularities} and,
although a number of achievements were recently made in this direction
\cite{n2}, a completely satisfactory picture of the gravity dual
is still missing.
The proposals involving maximally supersymmetric compactifications
of M-theory on $AdS_4\times S^7$
or $AdS_7\times S^4$  are instead limited by the poor
understanding of the superconformal field theories living on M-branes.
Despite these immediate difficulties, the correspondence have
enlightened a beautiful interplay between gauge theory and gravity
physics and the impressive amount of
results in the last few years justify the initial enthusiasm
 (for a review and references see \cite{review}).

The low energy physics around vacua of type II (or
11--dimensional) supergravities involving $AdS$ spaces times
spheres (or more general Einstein spaces) can be efficiently
described in terms of suitable gauged supergravities on $AdS$
vacua. These effective descriptions are believed to be consistent
truncations of a higher dimensional supergravity theory reduced on
the internal Einstein space. A solution of the gauged supergravity
defines a solution in the higher dimensional theory and
vice-versa, although the details of the lifting are often hard to
determine. In a nice series of works pioneered by \cite{fgpw},
domain wall solutions of five-dimensional ${\cal N}=8$
supergravity that interpolate between Anti-de Sitter vacua with
different number of supersymmetries were studied and a detailed
correspondence between bulk fields and composite operators in the
infrared gauge theory was constructed.
 Although the equations of
motion of gauged supergravities which describe interesting flows
are typically rather complicated to solve,
many important features of the flow
can already be read from the physics around the two fixed points.
Moreover the complete interpolating solution, when not known
analytically, can be dealt with numerically, obtaining
 some valuable information about the flow.
  In the cases where an analytic kink solution is available,
a more quantitative information like correlation functions,
scalar operator mixings, etc. can be determined explicitly from
the flow (see \cite{bfs} and references therein).

 The aim of the present paper is to provide examples of analytic
kink solutions of three dimensional gauged supergravity,
where the details of the flow and the lift
to nine or ten-dimensions can be explicitly displayed.
The solutions describe the near horizon limits of D1D5
and D1D5+KK monopoles bound state systems in
freely acting orbifold compactifications of type IIB.
They will always contain a trivial $T^4$ or $K3$ part on
which the D5 branes are wrapped. For simplicity we will omit
this part in most of our discussion and refer to the lift as a lift
to five or six dimensions.
The domain wall solutions will be determined as solution of
five-dimensional supergravity after reduction from more
familiar six-dimensional geometries on a circle with non-trivial
bounday conditions.
In the case of pure D1D5 system the solution
interpolates between an $AdS_3\times S^2$
and a dilatonic  $AdS_2\times S^3$  vacuum
of five dimensional supergravity.
The latter can be better described in terms
of a further lift to six-dimensions where it is given by
the more familiar $AdS_3\times S^3$ vacuum with constant dilaton.
In the case of D1D5+KK monopole
and fluxes, the five-dimensional solution can
be extended all the way out of throat to a Ricci-flat
asymptotic geometry with constant dilaton and therefore
a sensible five-dimensional description is avaible at the two ends
of the flow.

  From the CFT point of view the walls describe the RG-evolution
out of two dimensional ${\cal N}=(4,0)$ conformal field theories
living on the $AdS_3$ boundaries. Amazingly, the whole flow is
generated by a non-trivial choice of boundary conditions on the
familiar D1D5 systems for type IIB on $M_4$ with $M_4$ being $T^4$
or $K3$. More precisely the solution describes the near horizon
geometry of D1D5 systems in type IIB on $M_4\times
(\mathbb{R}^4\times S_1)/Z_N \sigma_{1\over N}$, with $Z_N$ acting
as a rotation of $\mathbb{R}^4$ and $\sigma_{1\over N}$ an order
$N$ shift along a longitudinal circle of radius $N R$. The two
fixed point geometries, which we will refer to as ``deep inside''
and ``asymptotic'' regions, can be reached by sending $R$ to zero
or infinity while keeping $N$ large but fixed. In the case
$R\rightarrow 0$ the effects of the shift can be neglected and the
system effectively lives in type IIB on $M_4\times S^1\times
\mathbb{R}^4/Z_N$ (see \cite{tu,dm} for example). At the other end
of the flow $R\rightarrow \infty$ the theory decompactifies to IIB
on $M_4\times \mathbb{R}^5$. Alternatively one can think of the
orbifold as a compactification of the system on a circle of radius
$R$, where fields on the circle are periodic only up to a $Z_N$
rotation of the transverse $S^3$. Upon reduction to five
dimensions this procedure leads to Melvin solutions with
non-trivial NS--NS fluxes.
 The $Z_N$ will be always embedded inside an $SU(2)_L$ subgroup
of the full $SO(4)$ isometry group of $S^3$ and therefore
the solution preserves half of the original ${\cal N}=(4,4)$
supersymmetries.
Similar ideas have been extensively exploited in the construction
of flux--brane solutions \cite{fluxbranes, gs} (see \cite{dggh} for
earlier works in the subject).
 The case of supersymmetric flux--branes have been first
discussed in \cite{gs}.
Supersymmetric versions of NSNS Melvin universes and
the spectra of open strings living on D-branes in these backgrounds
have been  discussed in \cite{tu}
(see also \cite{dm} for similar results in the context of RR flux--branes).

 The second part of the paper is devoted to the study of
the global properties of our solutions, seen as ``tilted''
locally $AdS_3\times S^3$ geometries in six dimensions. The
effects of the global identifications on $AdS_3$ geometries have
been studied in a beautiful paper by Brown and Henneaux \cite{bh}.
In particular,  they
showed how point mass solutions carrying
non--trivial momentum charges can
be constructed in asymptotically $AdS_3$
vacua by modding the geometry over global identifications.
 The effect of the orbifolding in our $AdS_3\times S^3$ geometry
is in some sense milder and the solutions still define vacua
carrying no charges at infinity.
In addition we show how  although the global
isometry group is drastically reduced from $SL(2,R)_L\times SL(2,R)_R\times SU(2)_L\times SU(2)_R$
to an $SU(2)_R\times U(1)^3$ subgroup, the full two-dimensional
conformal group is restored at infinity and is realized in terms of two
copies of a Virasoro algebra with the expected central charge.

The paper is organized as follows:
In section 2, we
construct, via NS--NS Melvin fluxes, domain wall
solutions of five-dimensional supergravity and discuss
their near and asymptotic fixed point limits.
Section 3 is devoted to the discussion the
global properties of the kink solutions
from the six-dimensional point of view.
In section 4 we include some concluding remarks and comment on
interesting directions of future research.

\section{Melvin universe as domain walls in five-dimensional supergravity}

In this section we
construct supergravity solutions corresponding
to geometries that look locally (but not globally) as products
of AdS spaces times spheres.
We follow closely the lines of \cite{fluxbranes} where similar
solutions were found for M--theory on Ricci flat spaces.
As in those cases the non--trivial warped geometry
descends from more familiar solutions in higher dimensions
upon reduction on a torus with unusual boundary conditions.
 More precisely
we consider the case where a loop in a compact coordinate
(not necessarily the eleventh coordinate)
is accompanied by a non--trivial shift on the transverse sphere.
Upon reduction to lower dimensions they lead to Melvin solutions
with non trivial fluxes and profiles for the dilaton field.
In the context of the AdS/CFT correspondence the isometries of
the spheres are related to R-symmetries of the boundary
conformal field theory. Turning on non-trivial fluxes will
then break part of these isometries leading
to less supersymmetric AdS/CFT duals.
We are interested in the case where the fluxes are chosen in
such a way that half of the original supersymmetries
are preserved.
A typical example of such a configuration is the flux 5-brane
\cite{gs} of type IIA, involving a reduction on the eleventh
dimensional circle accompanied by a $Z_N$
rotation in the transverse $R^4$.
More general solutions involving wrapped flux--branes
were studied in \cite{u}.

The general idea behind the construction of Melvin solutions
in Einstein gravity is quite simple and can be described as follows
\cite{melvin}. Consider a given solution of the Einstein-Hilbert equation of
motion in D-dimensions described by a metric  $G_{MN}$ (whose isometry
group contains the isometries of a d--dimensional torus $T_d$), a dilaton
profile $\phi$ and a set of non-trivial fluxes for the RR
rank $n$ field strength $H$. A solution is specified also
by a choice of boundary conditions along the directions of the torus.
Different choices lead to inequivalent
(sometimes drastically different)
physics  which share with the original solution only its
local characteristics.  After reduction to D-d dimensions they
give rise to a rich class of solutions of the lower
dimensional gravity with various non-trivial fluxes and scalar profiles.
More precisely, denoting the spacetime index by $\mu=0,...D-d-1$
and indicizing the directions of the torus $T_d$ by $i=D-d,..D-1$,
a general choice of boundary conditions is given by the identifications
\bea
x^i&\sim& x^i+ 2 \pi n^i\, R_i  \nn\\
x^\mu&\sim& x^\mu+2 \pi n^i\, b^{\mu}_i\, R_i
\label{bc0}
\eea
parameterized by the real parameters $b^{\mu}_i$, which describe
a jump of $2 \pi n^i \,b^{\mu}_i \,R_i$ along $x^\mu$ once
one goes $n^i$ times around the cycle ``i'' of $T^d$.
In order to perform a reduction to $D-d$ dimensions it is convenient
to introduce the coordinates $\tilde{x}^\mu\sim x^\mu- b^{\mu}_i\, x^i$
with the canonical orbits
\bea
x^i &\sim& x^i+ 2 \pi n^i\, R_i  \nn\\
\tilde{x}^\mu &\sim& \tilde{x}^\mu
\eea
In terms of these new coordinates the D-dimensional metric can be
rewritten (after reconstructing squares) as
\bea
ds_{D}^2&=&G_{MN}\, dx  \, dx^N \nn\\
    &=& g_{ij}\, (dx^i+A^i_\mu d\tilde{x}^\mu) \,
(dx^j+A^j_\mu d\tilde{x}^\mu)+
g_{\mu\nu}\, d\tilde{x}^\mu \, d\tilde{x}^\mu
\label{metricg}
\eea
with
\bea
g_{ij}&=&G_{ij}+ 2\, G_{\mu i}\, b^{\mu}_j+ G_{\mu\nu}
\, b^{\mu}_i\, b^{\nu}_j  \nn\\
A^i_{\mu}&=&g^{ij}\,\left( G_{\mu j}+\, G_{\mu\nu}\,b^{\nu}_j\right)\nn\\
g_{\mu\nu}&=&G_{\mu\nu}-A^i_{\mu}\,g_{ij}\, A^j_{\nu}
\label{fluxg}
\eea
After reduction on $T^d$ we are left with a ``dilatonic'' solution
with metric $g_{\mu\nu}$, non-trivial profiles $g_{ij}$
for the scalars coming from the metric and fluxes related to
 the lower dimensional gauge fields  $A^i_{\mu}$.
In addition the six-dimensional rank n field strength, which
in the new coordinates is given by
\be
H_{\tilde{M}_1....\tilde{M}_n}=
\partial_{\tilde{M}_1} x^{M_1}.....\partial_{\tilde{M}_n} x^{M_n}
\, H_{M_1....M_n}
\ee
gives rise to a rank $n-1$ and a rank $n$ forms given by
\bea
H^n_{\tilde{\mu}_1....\tilde{\mu}_{n-1}}&=&H_{\tilde{\mu}_1....\tilde{\mu}_{n-1} i}\nn\\
H^{n-1}_{\tilde{\mu}_1....\tilde{\mu}_n}&=&H_{\tilde{\mu}_1....\tilde{\mu}_n}-
H_{\tilde{\mu}_1....\tilde{\mu}_{n-1} i}\,
A^i_{\tilde{\mu}_n}+{\rm cyclic~~permutations}
\label{rrflu}
\eea
We will be mainly interested in the case $d=1$. Denoting
by ``$x$''  the compact coordinate $x_{D-1}$, the $D$-dimensional
metric can be rewritten as
\be
ds_{D}^2 = e^{2\sigma}\, (dx+A_\mu\, d\tilde{x}^\mu) \,
(dx+A_\nu\, d\tilde{x}^\nu)+
g_{\mu\nu}\, d\tilde{x}^\mu \, d\tilde{x}^\nu
\label{metric}
\ee
with $e^{2\sigma}\equiv \Lambda \,G_{xx}$ and $\Lambda$, the (D-1)-dimensional metric and
the gauge field potentials given by
\bea
\Lambda &=&  1 +2\, {G_{\mu x}\over G_{xx}}\, b^{\mu}+
{G_{\mu\nu}\over G_{xx}}\, b^{\mu}\, b^{\nu}\nn\\
A_{\mu}&=& e^{-2\sigma}\left( G_{\mu x}+ G_{\mu\nu}\, b^{\nu}\right) \nn\\
g_{\mu\nu}&=&G_{\mu\nu}-e^{2\sigma}\, A_{\mu}\, A_{\nu}
\label{flux}
\eea

Together with the $H$ fluxes (\ref{rrflu}), the field strength $F\equiv d A$,
the metric $g$ and the dilaton $e^{-2\phi}\equiv R\, e^{-2\Phi_D+\sigma}$,
they
define a solution of the (D-1)-dimensional equations of motion
coming from the supergravity theory with bosonic
action (in the string metric)
\bea
S&=&2\,\pi\, \int \, d^{D-1} x\, \sqrt{-g}\, \left[
e^{-2\phi}\left( {\cal R}+4\,(\partial \phi)^2-(\partial \sigma)^2-
{e^{2\sigma}\over 4} F^2\right)\right.\nn\\
&&\left.
-{e^\sigma \over 2 n!}\, H_n^2- {e^{-\sigma}\over 2 (n-1)!}\,
H_{n-1}^2 \right]
\label{action}
\eea

\subsection{D1D5 systems in presence of NS--NS fluxes}

In this subsection we construct flux solutions of five-dimensional
supergravity descending from $AdS_3\times S^3$ vacua.
The starting $AdS_3\times S^3$ geometry
 have been extensively studied
in the context of the AdS/CFT correspondence and are associated
to two-dimensional
CFTs describing the low energy excitations of bound states
of D1D5 branes (or a stack of NS5 branes and fundamental strings)
wrapping a four manifold M, with $M_4=T^4$ or $K3$.
It is natural to ask how different choices of boundary conditions
in the  ${\cal N}=(4,4)$ two-dimensional
CFT (after compactifying the six-dimensional
black string on a circle) are realized in the dual supergravity.

The near horizon solution describing a bound state of
$Q=Q_1 Q_5$ D1D5 branes
is described by the $AdS_3\times S_3$ six-dimensional metric
and self-dual RR field strength
\bea
ds^2&=&{r^2\over \ell^2}\,(-dt^2+dx^2)+{\ell^2\over r^2}\, dr^2
+\ell^2\, d \Omega_3\nn\\
H_{txr}&=&{2\,r\over \ell^2}\nn\\
H_{\theta\varphi_1 \varphi_2}&=&2\, \ell^2 \, \sin\, 2\, \theta
 \label{D1D5}
\eea
where $r^2=x_2^2+....x_5^2$ is
the radial distance from the D-brane system, $\ell^2=g_6\sqrt{Q}$ the
squared of the Anti-De Sitter and $S^3$ radius and
$g_6=g_{st}/\sqrt{v_M}$ the six-dimensional coupling constant
\footnote{We will
always measure distances in units of $\alpha^\prime$.}. Finally
\be
d \Omega_3 = d\theta^2 +d\varphi_1^2 +d\varphi_2^2 +
              2\, d\varphi_1\,d\varphi_2\,\cos \,2\, \theta,
\label{o3}
\ee
denotes the line element along the transverse three sphere with
$0\leq \theta \leq {\pi\over 2}$, $0\leq \varphi_1 \leq \pi$ and
$0\leq \varphi_2 \leq 2 \pi$.
 We would like to consider the effect of introducing non--trivial
boundary conditions along the ``$\sigma$'' direction
on the worldvolume CFT, which is set to coincide with the
``$x$'' direction on the boundary of $AdS_3$.
 In order to preserve supersymmetry we consider identifications
like (\ref{bc0}), where
the spacetime shift is embedded on a $SU(2)_L$ subgroup of the full
$SU(2)_L\times SU(2)_R$ isometry group of $S^3$ .
More precisely we consider the case where fields are
taken to be periodic on $x$ only up to a $Z_N$ rotation
of the transverse $R^4$ \cite{gs}
\bea
x&\sim& x+ 2 \,\pi\, n\, R  \nn\\
\varphi_2&\sim& \varphi_2+2\, \pi\, n\, b\, R
\label{bcD1D5}
\eea
with $b\, R={1\over N}$. Clearly this orbifolding preserves only
an ${\cal N}=(4,0)$ subset of the original two-dimensional
supersymmetries with R-symmetry group now reduced to
$SU(2)_R\times U(1)_L$.
From the five dimensional point of view (after reduction on $x$)
our new supergravity solution can be read off from (\ref{metric},\ref{flux})
and is described by the metric
\bea
ds_5^2 &=& -{r^2\over \ell^2}\,dt^2+{\ell^2\over r^2}\, dr^2
+\ell^2\, d\theta^2\nn\\
&& +{\ell^2\over \Lambda\,}
\left[ d\tilde{\varphi}_2^2 + 2\,\cos \theta\, d\varphi_1\,d\tilde{\varphi}_2
+ d\varphi_1^2\,(1+ {b^2\over r^2}\,\ell^4\,sin^2\,2\, \theta)\right]
\label{metricD1D5}
\eea
in terms of the new variable
$\tilde{\varphi}_2\equiv \varphi_2-b \, x$. The remaining
five dimensional fields are given by
\bea
\Lambda&=& 1+ {b^2 \,\ell^4\over r^2}\nn\\
e^{-2(\phi-\phi_\infty)}&=& R\, e^\sigma=R\,{r\over \ell}\, \Lambda^{1\over 2}\nn\\
A_{\varphi_1} &=& {b\, \ell^4\, \cos \,2\, \theta \over \Lambda\, r^2}\nn \\
A_{\tilde{\varphi}_2}&=&{b\, \ell^4 \over \Lambda\, r^2}\nn\\
H_{tr}&=&  {2 \,r\over \ell^2} \nn\\
H_{\theta\varphi_1}&=&2\,b\, \ell^2 \, \sin\, 2\, \theta\nn\\
H_{t r \mu}&=&- {2\,r \over \ell^2}\, A_\mu \nn\\
H_{\theta\varphi_1\tilde{\varphi}_2}&=& {2\,\ell^2\over \Lambda}
\, \sin\, 2\, \theta \label{fluxD1D5} \eea Besides the background
fields descending from the original RR sources we see from
(\ref{fluxD1D5}) that the new solution involves non-trivial NS--NS
fluxes $A_\mu$ for the gauge field $g_{\mu x}$. These fluxes are
responsible for the partial supersymmetry breaking and the
consequent non-trivial profile of the dilaton field and the
metric. In the next section we will present a detailed study of
the global features of this new supergravity solution from the six
dimensional point of view, but before that we would like to see
how much we can learn from the local physics described by the
metric and background fields above. In particular it is
interesting to notice that the flux parameter $b$ sets a new scale
in the near horizon geometry. Indeed
(\ref{metricD1D5},\ref{fluxD1D5}) can be seen as a domain wall
solution interpolating between  regions of small ($r\ll b \ell^2$)
and large ($r\gg b \ell^2$) radial distances. Remarkably both
regimes can still be accurately described inside perturbation
theory and  we will refer to them as the ``deep inside '' and
``asymptotic'' regions respectively. Let us first consider the
solution in the deep inside region $r\ll b \ell^2$. The limit can
be achieved by turning the flux parameter $b$ to infinity while
keeping  $b\,R=1/N$ fixed and small, much in the same way as the
near horizon geometries can be recovered from large N expansions
of the exact supergravity solutions. To this end it is convenient
to introduce the rescaled coordinates $\hat{\varphi_2}\equiv
{\tilde{\varphi}_2\over b}$, $\hat{\varphi_1}\equiv 2\,\varphi_1 $
and $\hat{\theta}\equiv 2\, \theta $. The metric
(\ref{metricD1D5}) may then be written (up to orders ${1\over b}$)
as \bea ds_{\rm near}^2={r^2\over
\ell^2}\,(-dt^2+d\hat{\varphi}_2^2) +{\ell^2\over r^2}\, dr^2
+{\ell^2\over 4}\,(d\hat{\theta}^2+ \sin^2\,\hat{\theta}\,
d\hat{\varphi}_1^2) \label{ads3s2} \eea which describes an
$AdS_3\times S^2$ space with radius $\ell$ and $\ell/2$
respectively. In a similar way one can evaluate the limit of large
$b$ for the remaining NS--NS/RR backgrounds in (\ref{fluxD1D5}).
The surviving components are given by \bea
e^{-2(\phi-\phi_\infty)}&=&R\, e^\sigma= b \,R\, \ell={1\over N}\, \ell  \nn\\
e^\sigma\, A_{\hat{\varphi}_1} &=& {1\over 2}\,\ell \, \cos \, \hat{\theta}
\nn\\
e^{-\sigma}\,H_{\hat{\theta}\hat{\varphi}_1} &=&
{1\over 2}\, \ell \, \sin\, \hat{\theta}\nn\\
H_{t r \hat{\varphi}_2}&=&- {2\, r \over \ell^2}
\label{fluxD1D5n}
\eea
It is worth stressing that the metric (\ref{ads3s2}) and
background fields (\ref{fluxD1D5n})
define a solution of the Einstein--Hilbert equations of motion by itself
and can therefore be extended to any $r$. \par
 Notice that in the limit $b\rightarrow \infty$, the space direction on the boundary of $AdS_3$, parameterized by $\hat{\varphi}_2$,
 ``leans'' towards the $x$-direction in the original$AdS_3$ boundary.
Indeed the deep inside solution (\ref{fluxD1D5n}, \ref{ads3s2})
can be alternatively derived from reduction on the $\varphi_2$
fiber inside $AdS_3\times S^3/Z_N$\footnote{Throughout the text we
keep $N$ fixed but large.}. This is in agreement with the
expectations for the near horizon geometry of the D1D5 system on
$M_4\times S^1\times\mathbb{R}^4/Z_N$. The above analysis can be
extended to the whole flow by exchanging the role of $x$ and
$\varphi_2$ in (\ref{bcD1D5}), and rewriting all five dimensional
quantities in terms of $\tilde{x}=x-{\varphi_2\over b}$. We will
not present here the details of this equivalent description of the
flow, which follows similar lines as that presented above and
leads to identical conclusions.

Turning the flux parameter $b$ to zero one can in a similar way isolate the
asymptotic geometry and background fields. Notice that in this limit
the five dimensional dilaton is no longer  constant and the solution is better described in
six-dimensional terms where we recover our starting $AdS_3\times S^3$
metric and background fields (\ref{D1D5}). The boundary of $AdS_3$
decompactifies in this case to Minkowski $M_{1,1}$ since
$R$ should be consistently taken to infinity in this limit in
order to keep $b\, R={1\over N}$ finite.

\subsection{Adding KK monopoles }

The D1D5 solution with fluxes (\ref{D1D5}) smoothly interpolates
between a nearby and an asymptotic $AdS_3$ geometries naturally
living in five and six
dimensions respectively. The effective size of the sixth dimension is related
to the five-dimensional dilaton and grows to
infinity far away from the brane. This asymptotic behavior,
although shared by most of the flux-brane configurations studied
in the literature,
is not generic to  all known Melvin universes.
In \cite{u}, the author shows how
in the presence of Taub-Nut geometries, flux branes get trapped and
the region where the lower dimensional picture
breaks down (since dilaton diverges), is cut off.
We would like now to exploit this idea to construct
an interpolating solution where both ends of "the flow"
admit a sensible five-dimensional interpretation.

The solution will be associated with non-trivial boundary conditions for
a bound state system of D1D5 branes
and KK monopoles in type IIB.
In the absence of fluxes the near horizon geometry of this system
can be obtained from the $AdS_3\times S^3$ metric describing
the pure D1D5 system by replacing the $R^4$ cone
over $S^3$ by a Taub-Nut space.
For $Q_k$ coinciding KK monopoles this recipe yields a supergravity
background with $AdS_3\times S^3/Z_{Q_k}$ geometry, which is
believed to be holographically
dual to the ${\cal N}=(4,0)$ boundary CFT describing the excitations
of a bound state system of D1D5 branes and KK monopoles \cite{bps}.
A dual version of this
correspondence has been
extensively studied in \cite{klr}.

In the presence of fluxes the analysis of the near horizon geometry

follows closely
our previous results but the two geometries differ drastically
in the asymptotically far regime. Another important difference
with our former example is that the fluxes do not break
additional supersymmetries among those already preserved by
the D1D5KK system (this holds true already in the absence of D-branes, see
\cite{u}).

The starting metric (see for instance \cite{tsetse} and references therein)
reads:
\bea
ds^2 &=& H^{-1}(-dt^2 +dx^2)+ H\,
\left[H_k^{-1}\,
 (d\tau + Qk\,(1- \cos \, \theta) \, d\phi)^2\right. \nn\\
&&+ \left. H_k \, ( dr^2 + r^2\, d\theta^2 +
r^2\, d\phi^2\, \sin^2 \theta) \right]
\eea
with
\bea
H=1+{\ell^2\over r}\quad\quad
H_k=1+{Q_k\over r}
\eea
the harmonic functions associated with the $\ell^2=g_6\sqrt{Q_1 Q_5}$
branes and KK monopole
charges. In addition the D1D5 background include the
self-dual RR field
strength
\bea
H_{txr}&=& \partial_r H^{-1}\nn\\
H_{\theta\phi \tau}&=&  \ell^2  \, \sin\, \theta
\eea
After the identifications
\bea
x&\sim& x+ 2 \pi n\, R  \nn\\
\tau &\sim& \tau+4 \pi n\, b\, R
\eea
and reduction on $x$ we are left with the five dimensional metric
(in terms of the new variable $\tilde{\tau}=\tau-2\,b\, x$.
\bea
ds_5^2 &=& -H^{-1}\, dt^2+ H \,H_k( dr^2 + r^2\, d\theta^2+r^2\,\sin^2 \theta\,d\phi^2)\nn\\
&&+{H \, H_k^{-1}\over \Lambda}\,
(d\tilde{\tau} + Qk\,(1- \cos \, \theta) \, d\phi)^2
\label{metricD1D5kk}
\eea
and NS--NS/RR fields
\bea
\Lambda &\equiv& 1+4\,b^2\, \, H^2 \, H_k^{-1} \nn\\
e^{-2\phi}&=& R\,e^{\sigma}=  R\,H^{-{1\over 2}}\,\Lambda^{1\over 2}\nn\\
A_\phi&=& {2\,b H^2 Q_k (1- \cos \, \theta)\over \Lambda\, H_k} \nn\\
A_{\tilde\tau}&=&{2\,b H^2  \over \Lambda\, H_k} \nn\\
H_{tr}&=& \partial_r H^{-1}\nn\\
H_{\theta\phi}&=&  2\,b\, \ell^2 \, \sin\, \theta \nn\\
H_{\theta\phi\tilde{\tau}}&=&  { \ell^2 \, \sin\, \theta\over \Lambda}\nn\\
H_{tr\mu}&=& -\partial_r H^{-1}\, A_\mu
\label{fluxD1D5kk}
\eea
 The two interesting ``fixed point'' geometries are
now recovered in the limits $r\ll {b^2 \ell^4\over Q_k}$
and $r\gg {b^2 \ell^4\over Q_k}$.
The crucial difference with our previously studied
example is that now, in both the limits $r\rightarrow 0$
or $r\rightarrow \infty$, the dilaton $e^\sigma$
stabilizes leading to a solution with sensible five dimensional
description.
 Indeed the solution interpolates between an
asymptotically far solution with Ricci flat metric
$R\times {\rm \left( Taub-Nut\right)}_{\infty}$
with trivial background fields and
the deep inside $AdS_3\times S^2$ geometry
\bea
ds_{\rm near}^2={r \over
\ell^2}\,(-dt^2+d\hat{\tau}^2)
+{\ell^2\, Q_k \over r^2}\, dr^2
+\ell^2\, Q_k \,(d\theta^2+ \sin^2\,\theta\, d\phi^2)
\label{ads3s2KK}
\eea
with $\hat{\tau}={\tau\over 2\,b}$ and radii
 $2\, \ell\, \sqrt{Q_k}$ and $\ell\, \sqrt{Q_k}$
for the two pieces respectively.
The surviving background fields in this limit are given by
\bea
e^{-2\phi}&=&R\, e^\sigma={2\, b \,R\,\ell\over \sqrt{Q_k}}=
{2\,\ell\over N\,\sqrt{Q_k}}
 \nn\\
e^\sigma\, A_{\phi} &=&\ell\, \sqrt{ Q_k}\, (1-\cos \, \theta)\nn\\
e^{-\sigma}\,H_{\theta\phi} &=& \ell\,\sqrt{ Q_k} \, \sin\, \theta\nn\\
H_{t r \hat{\tau}}&=& {1 \over \ell^2}
\label{fluxD1D5KKn}
\eea

\section{Study of global properties of the solutions: charges at infinity}

In the previous sections we have constructed new supergravity
solutions that look locally (but not globally) like products of
$AdS$ spaces times spheres. The aim of this section is to present
a systematic study of the global properties and charges
characterizing these brane geometries. We will adopt the
Hamiltonian formalism in General Relativity \cite{Regge, Wald},
where charges associated with asymptotic isometries of a given
space--time solution are unambiguously defined.

\subsection{Hamiltonian formalism in
General Relativity}

Let us briefly review the Hamiltonian formalism in General
Relativity \cite{Regge, Wald} and introduce the basic definitions
and notations that will be extensively used in this section. The
first step is to formalize the concept of time evolution of a
system. To this end one introduces a  globally defined vector
field $t^a$ and a function $t(x)$ such that $t^a\,\nabla_a t=1$.
The loci of constant $t(x)$ are space--like hyper--surfaces
denoted by $\Sigma_t$ while the vector $t^a$ is chosen to define
the time evolution of the quantities restricted to $\Sigma_t$ (at
least locally one would be able to define a time coordinate $t$
and $n-1$ space coordinates $x^i$ such that
$t^a=(\partial/\partial t)^a$). We also choose a volume form
$\epsilon^{(n-1)}_{a_1\,\dots\,a_{n-1}}=
\epsilon_{a\,a_1\,\dots\,a_{n-1}}\,t^a$ for $\Sigma_t$ which is
invariant under time evolution (i.e. diffeomorphism generated by
$t^a$): $\mathcal{L}_t\,(\epsilon^{(n-1)})=0 $. Finally we define
our coordinate system such that $\epsilon^{(n-1)}$ has non
vanishing components $\pm 1$.\par

The definition of a space--like hyper--surface and a time
direction allows to introduce canonical variables which define the
phase space of the system. In this formalism the generator of the
time evolution, i.e. the Hamiltonian, will be denoted by $H[t^a]$.
If we consider the pure gravity case, the system is totally
described by the metric $g_{a b}$. We may express $g_{a b}$ in
terms of the induced metric ($h_{a b}$) on $\Sigma_t$ and of its
components out of $\Sigma_t$ given in terms of the extrinsic
curvature $K_{ab}$. The {\bf induced metric} $h_{a b}$ is defined
by means of a unit time--like vector $n^a$ (but not necessarily a
geodesic) orthogonal
 to $\Sigma_t$:
\begin{eqnarray}
h_{a b}&=&g_{a b}+n_a\,n_b\label{hab}
\end{eqnarray}
${h_a}^b$ being simply the projector on $T(\Sigma_t)$. The {\bf
extrinsic curvature} $K_{a b}$ is defined on the other hand as the
gradient of $n^a$ along $\Sigma_t$
\begin{eqnarray}
K_{a\,b}&=&{h_a}^c\nabla_c\,n_b \label{kab2}
\end{eqnarray}
Finally we introduce a covariant derivative $D_a$
which acts on tensor fields ${T_{a_1\dots a_r}}^{b_1\dots b_s}$
on $\Sigma_t$ as:
\begin{eqnarray}
D_a\,{T_{a_1\dots a_r}}^{b_1\dots b_s}&=&{h_a}^c\,{h_{a_1}}^{c_1}\,
\dots {h_{a_r}}^{c_r}\,{h^{b_1}}_{d_1}\,\dots {h^{b_s}}_{d_s}\nabla_c\,
{T_{c_1\dots c_r}}^{d_1\dots d_s}
\end{eqnarray}

We are now ready to define the canonical variables and the Hamiltonian $H\equiv H[t^a]$.
The Einstein Lagrangian can be rewritten in terms of quantities
related to $\Sigma_t$ as:
\begin{eqnarray}
{\cal L}_G &=&\sqrt{-g}\,R[g]=t_\bot \,\sqrt{h}\,\left({\cal R}
+({K_a}^a)^2-(K_{ab}\,K^{ab})\right)
\end{eqnarray}
where we denote by $t_\bot,\, t_{\|}^a$ the directions of $t^a$
parallel and orthogonal to $\Sigma_t$
\begin{eqnarray}
t^a &=& t_\bot \,n^a+t_{\|}^a
\label{tnN}
\end{eqnarray}
The momentum $\pi^{ab}$ conjugate to the field $h_{ab}$ is defined
in the following way: \be \pi^{ab}=\frac{\delta {\cal L}_G}{\delta
\dot{h}_{ab} }=\sqrt{h}\,\left(K_{ab}-h_{ab}\,K\right) \ee with
$\dot{h}_{ab}=
{h_a}^{a_1}\,{h_b}^{b_1}\,\mathcal{L}_t\,h_{a_1b_1}$. Notice that
the components of $t^a$ do not appear in the Lagrangian through
time derivatives and therefore
 they are not dynamical variables and have no associated
conjugate momentum.

The Hamiltonian $H$ is expressed as:
\bea
H&=&\int_{\Sigma_t}\,\epsilon^{(n-1)}\,\left(\pi^{ab}\,\dot{h}_{ab}-{\cal
L}_G\right)+ J[t^a] =\int_{\Sigma_t}\,\epsilon^{(n-1)}\,\left(
t_\bot\,{\cal H}_\bot+t^a\,{\cal H}_{\|\,a}\right) + J[t^a] \nn \\
{\cal H}_\bot &\equiv&-{\cal R}+\frac{1}{h}\left[\pi_{ab}\,
\pi^{ab}+\frac{\pi^2}{(2-n)}\right]\nonumber\\
{{\cal H}_{\|}}^{a}&=&-2\, D_b\,\pi^{ab} \label{gen1} \eea The
additional boundary term $J[t^a]$  was shown in \cite{Regge} to be
required in order for the functional derivatives of H with respect
to the canonical variables to be well defined, namely in order for
them to vanish on $\partial \Sigma_t$, so that the boundary
conditions on $h_{ab}$ and $\pi^{ab}$ at spatial infinity are not
affected by time--evolution. This quantity provides a definition
of the global charge associated with $t^a$ on the solution, which
is the total energy of the configuration.
 From functional derivation of $ H$
with respect to $t^a$ we deduce two phase space constraints ${\cal
H}_\bot={\cal H}_{\|\,a}=0$.

 In analogy with the definition of $H[t^a]$ as the generator of time
evolution, given an asymptotically Killing vector $\xi\in T(M_n)$
we may define the corresponding charge $H[\xi]$ as the following
generator in the phase space: \be
H[\xi]=\int_{\Sigma_t}\,\epsilon^{(n-1)}\,\left(\xi_\bot\,{\cal
H}_\bot+\xi^a_{\|}\,{\cal H}_{\|\,a}\right)+J[\xi] \label{Q} \ee
 where $J[\xi]$ is the boundary contribution analogous to $J[t^a]$ for the
 energy. An
explicit expression for the functional variation of $J[\xi]$ was
derived in \cite{bh} and reads {\small
\begin{eqnarray} \delta J[\xi]&=&\oint_{\partial\Sigma_t}\,dS_d
\,\left[ G^{abcd}\,
\left(\xi_\bot\,D_c-\partial_c\xi_\bot\right)\,\delta
h_{ab}\right. \nn\\
&& \left. +\left(2\,\xi_{\|}^b\,\pi^{ad}-\xi_{\|}^d\,\pi^{ab} \right)\,\delta
h_{ab}+2\,\xi_{\| a}\,\delta\pi^{ad}\right]
\label{dJ}
\end{eqnarray}}
with  $G^{abcd}=(\sqrt{h}/2)
(h^{ac}\,h^{bd}+h^{ad}\,h^{bc}-2\,h^{ab}\,h^{cd})$. For solutions
of the Einstein equation of motion we have ${\cal H}_\bot={\cal
H}_{\|\,a}=0$ and therefore $J[\xi]$ represents the only
contribution to the charge (\ref{Q}). For asymptotically flat
space-times one can easily see that this general expression
reduces to the standard definitions of the energy and momentum
charges:
\begin{eqnarray}
J[\partial_t]&=&
\lim_{r\rightarrow \infty}\,\oint_{\partial\Sigma_t}\,dS_k
\,\sqrt{h}\,\left(
\partial^i\, {h_{i}}^k-\partial^k\, {h_{i}}^i\right)~~~{\rm ADM~ mass}\nn\\
J[\partial_i]&=&\lim_{r\rightarrow \infty}\,2\,\oint_{\partial\Sigma_t}\,
dS_k\, \pi^{ik}~~~~~~~~~~~~~~~~{\rm Momentum~ along}~x^i
\end{eqnarray}
In the next subsection we will evaluate, using the general
expression (\ref{dJ}), the global charges
 characterizing the various locally AdS geometries
previously defined.

\subsection{Asymptotic $AdS_3\times S_3$ isometries and central charge.}

In this section we study the global properties of the
six-dimensional D1D5 geometry (\ref{D1D5}) with boundary
conditions (\ref{bcD1D5}). The non--trivial identifications in
$x,\,\varphi_2$ break the global $AdS_3\times S_3$
isometry group down to a $U(1)^3\times SU(2)_R$ subgroup with
Cartan generators
$\partial_x,\partial_t,\partial_{\varphi_2},\partial_{\varphi_1}$.
With each of these Killing vectors we can associate a global
charge through (\ref{dJ}). More generally, a conserved charge can
be associated with each asymptotic (not necessary global) Killing
isometry. We will see how, even in the presence of non-trivial
boundary conditions ( $b\neq 0$ in (\ref{bcD1D5})) a full 2d
conformal group is realized in the asymptotically far geometry in
terms of two copies of the Virasoro algebra with the expected
central charge.

The $AdS_3\times S_3$ metric (\ref{bcD1D5}), after the global
identifications (\ref{bcD1D5}), can be written as
\begin{eqnarray}
ds^2&=&\frac{r^2}{\ell^2}\left(-dt^2+dx^2\right)
+\ell^2\, \left[\frac{dr^2}{r^2}+  d\theta^2 +
d\varphi_1^2 +(d \tilde{\varphi}_2+b\, dx)^2\right. \nn\\
  &&\left.  + 2\, d\varphi_1\,(d\tilde{\varphi}_2+b \, dx)\,\cos \,2\, \theta
\right]
\label{metb}
\end{eqnarray}
in terms of the variables $x\sim x+2\, \pi\, R$,
 $\tilde{\varphi}_2\sim \tilde{\varphi}_2+2\, \pi$ with
standard orbits. The metric (\ref{metb}) is preserved
by the Killing generators
 \begin{eqnarray}
J_R^{(1)}&=&\cos ({2 \,\varphi_1})\,\partial_\theta- \cot (2\,\theta )\,
\sin ({2 \,\varphi_1})\,\partial_{\varphi_1}+\csc (2\,\theta
)\,\sin ({2 \,\varphi_1})\,\partial_{\tilde{\varphi}_2}\nonumber\\
J_R^{(2)}&=&-\sin ({2 \,\varphi_1})\,\partial_\theta-
 \cot (2\,\theta )\,\cos ({2 \,\varphi_1})\,\partial_{\varphi_1}
+\csc (2\,\theta
)\,
\cos ({2 \,\varphi_1})\,\partial_{\tilde{\varphi}_2}\nonumber\\
J_R^{(3)}&=&\partial_{\varphi_1}\nonumber\\
J_L^{(3)}&=&\partial_{\tilde{\varphi}_2}\nn\\
L_0 &=&{i\, R \over 2}\, \left( \partial_t-\partial_x+
b\, \partial_{\tilde{\varphi}_2}\right)\nn\\
\bar{L}_0 &=&{i\, R \over 2}\, \left( \partial_t+\partial_x-
b\, \partial_{\tilde{\varphi}_2}\right)
\label{siglobal}
\end{eqnarray}
which correspond to the global $ U(1)^3\times SU(2)_R\subset
SL(2,\mathbb{R})_L\times SL(2,\mathbb{R})_R\times SU(2)_L\times
SU(2)_R$ left unbroken by non-trivial identifications. A closer
look into the Killing equations derived from the metric
(\ref{metb}) reveals however that a richer isometry algebra is
restored in the asymptotically far region $r\rightarrow \infty$.
The group of asymptotic isometries can indeed be identified with
the full two-dimensional conformal group like in the more familiar
global $AdS_3\times S^3$ instance. The new asymptotic Killing
generators, realize two copies of the Virasoro algebra and can be
written as {\small \bea \label{asymptotic}
  L_n&=&\frac{i \,R}{2}\, e^{ i\,n\,(t-x)/R}\,
\left[ \left(1 - {n^2\,\ell^4\over 2\,R^2\, r^2} \right)
  \,\partial_t -\left(1 + {n^2\,\ell^4\over 2\,R^2\, r^2} \right)
  \,(\partial_x-b \, \partial_{\tilde{\varphi}_2}) -\frac{i\,n\, r}{R}\partial_r
\right] \nn\\
  \bar{L}_n&=&\frac{i\, R}{2} \,e^{ i\,n\,(t+x)/R}\,
\left[ \left(1 - {n^2\,\ell^4\over 2\,R^2\, r^2} \right)
  \,\partial_t +\left(1 + {n^2\,\ell^4\over 2\,R^2\, r^2} \right)
  \,(\partial_x-b \, \partial_{\tilde{\varphi}_2}) -\frac{i\,n\, r}{R}\partial_r
\right] \label{ln} \eea} The vector fields above generate
asymptotic Killing isometries in the sense that all non-trivial
variations of the metric (\ref{metb}) generated by them: \bea
\delta_n\, g_{\alpha\beta}&=&\nabla_{(\alpha}\, L_{\alpha)n}=
-\nabla_{( x} L_{t) n}=
-{n^3\,\ell^2\over 2\,R^2} e^{ i\,n\,(t-x)/R} \nn\\
\delta_{\bar{n}}\, g_{\alpha\beta}&=&\nabla_{(\alpha }\,
\bar{L}_{\beta)n}= -{n^3\,\ell^2\over 2\,R^2} e^{
i\,n\,(t+x)/R}\quad ;\quad \alpha,\beta=\{t,\,x \} \eea fall off
fast enough to leave unaltered the leading behavior of
(\ref{metb}) at space infinity.

As before, with each generator  (\ref{asymptotic}) of the
asymptotic isometries we associate a charge through (\ref{Q}). We
consider the most general linear combination of the Cartan
generators $\xi=a_1\, \partial_{t}+ a_2\,\partial_{\phi}+
a_3\,\partial_{\varphi_1}+ a_4\,\partial_{\tilde{\varphi}_2}$, and
read the corresponding charge from the $a_i$ coefficients of the
final answer. Charges associated with non Cartan generators are
vanishing since they always involve trivial integrals over sines
or cosines over $\partial_t \equiv S_1$.

The computation of the charge is simplified by the fact that the
extrinsic curvature of the hyper--surface  $\Sigma_t$ vanishes and
therefore only the first two terms in (\ref{dJ}), which do not
involve the canonical momenta $\pi^{ab}$ or its variations,
contribute to the charge. Indeed, being the metric t-independent
and block diagonal, the induced metric on $\Sigma_t$ is simply the
restriction $h_{ij}=g_{ij}$ with indices $i,j\neq t$ of $g_{ab}$
along $\Sigma_t$. In addition the only non-trivial component of
the covariant derivative of
 ${\bf n}={\ell\over r}\partial_t$ is $\nabla_r\,{\bf n}^t$
which does not have components along $\Sigma_t$ and therefore
yields a vanishing  extrinsic curvature according to the
definition (\ref{kab2}).
   We are left with potential contributions coming only
from the terms
involving  $\xi_\bot=-{a_1 r\over L}$ in (\ref{dJ}).
 These two contributions can be easily worked out and vanish
identically. In order to see this, let us recall that \be \delta
\, h_{cd}={\partial\, h_{cd}\over \partial b}\,db \ee have
components along the $(x,\varphi_1,\tilde{\varphi}_2)$ plane. On
the other hand the indices ``d'' and ``c'' in eq. (\ref{dJ}),
referring to the direction normal to the boundary
$\partial\Sigma_t$ at $r=\infty$ and to the components of the
derivatives respectively, can only be ``r''. Once again using the
fact that the metric is block diagonal (with respect to
$\partial\Sigma_t$) the only contribution to (\ref{dJ}) can only
come from the term proportional to $h^{ab}\delta\,h_{ab}$ which
clearly vanishes. We conclude that our vacuum configuration
carries vanishing charges with respect to all Killing isometries.

Even if all Virasoro charges are zero, once evaluated on our
background one can still compute the central extension by properly
evaluating the boundary contributions (\ref{dJ}) to the relevant
commutators $\left[ L_n,\,L_{-n} \right]$. We will however
postpone this computation to the next section, where a deformation
of our solution carrying non-trivial $L_0$ and $\bar{L}_0$ charges
are constructed and both terms in the Virasoro algebra (see eq.
(\ref{Virasoro}) below) can be displayed.

\subsection{A point mass solution}

In this subsection we follow the strategy of \cite{bh} in order
to construct  more general solutions
carrying non-trivial mass and momentum charges.
We introduce a new variable
$\phi=x/\ell$ and consider the metric:
\begin{eqnarray}
ds^2&=&-\left(1+\frac{r^2}{\ell^2}\right)\, dt^2+r^2\, d\phi^2+
\frac{dr^2}{\left(1+\frac{r^2}{\ell^2}\right)}+\nonumber\\
&&\ell^2\,\left( { {d\theta }}^2 + { {d\varphi_1}}^2 +
 {(d\tilde{\varphi}_2+b\,\ell\,d\phi)}^2 +
2\, {d\varphi_1}\, {(d\tilde{\varphi}_2+
b\,\ell\,d\phi)}\, \cos (2\,\theta ) \right)
\label{metbnew}
\end{eqnarray}
which clearly becomes (\ref{metb}) for $r\gg \ell$. In addition
one can easily verify that the Ricci tensors associated with the
two metrics (\ref{metb}) and (\ref{metbnew}) coincide and
therefore they are solutions of the same Einstein equations
\footnote{ Notice that in absence of flux ,i.e. $b=0$, and in the
limit of $R\rightarrow \ell$, the metrics (\ref{metb}) and
(\ref{metbnew}) are related by a change of coordinates and
correspond  to global $AdS_3\times S^3$.
 Solutions displaying asymptotic  $AdS_3\times S^3$ geometry
and carrying non trivial global charges have been recently
studied in \cite{bdkr}.}.
 The periodicity of
$\phi$ however is: $\phi\sim \phi+2\pi \alpha$ with $\alpha=R/ \ell$
and therefore whenever $\alpha\neq 1$ we encounter a conical singularity
at the origin. This corresponds to a point mass particle sitting at
the origin with a non-trivial contribution to the energy.

 More generally, following Brown and Henneaux
we can accompany the cycle around
$\phi$ by a jump in time:
\begin{eqnarray}
t&\sim & t-2\pi\,n\, A\nonumber\\
\phi &\sim & \phi+2\pi\,n\, \alpha
\end{eqnarray}
This identification will generate a momentum charge.
A convenient choice of coordinates is given by the replacements
\begin{eqnarray}
t&\rightarrow &\alpha\, t-A\, \phi\nn\\
\phi&\rightarrow & -\frac{A}{\ell^2}\, t+\alpha\, \phi\nonumber\\
r&\rightarrow&\frac{r}{\sqrt{\alpha^2-\frac{A^2}{\ell^2}}}
\end{eqnarray}
which give back to $\phi$ the standard period and removes the jump
in time:
\begin{eqnarray}
\phi&=& \phi+2\pi n\nn\\
t&\sim & t
\end{eqnarray}
The metric in the new variables reads
\begin{eqnarray}
ds^2 &=&-\left({r^2\over \ell^2}-A^2\, b^2+\alpha^2\right)\,dt^2+
\left(r^2-A^2+ b^2\,\ell^4\, \alpha^2\right)\,d\phi^2
+A\,\alpha\,(1-b^2\, \ell^2)\, d\phi\, dt\nn\\
&& +\ell^2 \, d\Omega_3+\frac{\ell^2\,dr^2}{(r^2-A^2+\ell^2 \alpha^2)}+
2\,b\,\ell\,( \ell^2\,\alpha\,d\phi-A\, dt)(d\tilde{\varphi}_2+
\cos{2\theta}\,d\varphi_1)
\label{finalmetric}
\end{eqnarray}
which tends to $AdS_3\times S^3$ at infinity.

We are now ready to compute the energy-momentum charges carried by
the solution (\ref{finalmetric}).
As before the charges are evaluated through the
boundary integral (\ref{dJ}).
The orthonormal vector ${\bf n}$ to the hypersurface
$\Sigma_t$, the induced metric $h_{ab}$ and
the canonical momentum $\pi^{ab}$ are now given by
 \begin{eqnarray}
{\bf n} &=& {1\over r}\,\left( \ell\, \partial_t -
{A\, \ell\, \alpha\over r^2}\, \partial_\phi+ A\, b\, \partial_{{\varphi}_2}
\right)\nn\\
h_{ab}&=& -A^2\,b^2\, dt^2+ d\phi^2 \, r^2 +
2\,A\,\alpha\,\left( -1 + b^2\, \ell^2 \right) {dt}\, {d\phi } +
 \frac{\ell^2}{r^2}\, dr^2+ \ell^2 d\Omega_3\nn\\
&& +2\, b\,\ell\,( \ell^2\,\alpha\,d\phi-A\, dt)(d\tilde{\varphi}_2+
\cos{2\theta}\,d\varphi_1)+...
\nonumber\\
\pi^{r\phi}&=&\pi^{\phi r}=
- \frac{A\,\alpha \,\ell^3\,\sin{2\,\theta}}{r^2}+... \nonumber\\
\pi^{r\varphi_2}&=&\pi^{\varphi_2 r}=
 \frac{A\,b\,\alpha^2 \,\ell^3\,\sin{2\,\theta}}{r^2}+...
\label{hn2}
\end{eqnarray}
As before we compute the charge associated with the general linear
combination $\xi=a_1\, \partial_{t}+ a_2\,\partial_{\phi}+
a_3\,\partial_{\varphi_1}+ a_4\,\partial_{\tilde{\varphi}_2}$, and
read the corresponding charge from the $a_i$ coefficients of the
final answer. The decomposition into normal and parallel part
leads to:
\begin{eqnarray}
\xi_{\bot}&=& {a_1\,r\over \ell}\nn\\
\xi_{||}^a &=& a_2\, \partial_\phi+a_3\partial_{\varphi_1}
+(a_4-{A\, a_1\, b\over \ell})\,\partial_{\tilde{\varphi}_2}
\label{xinp}
\end{eqnarray}
Plugging (\ref{hn2},\ref{xinp}) in (\ref{dJ}) one finds that only
the energy and momentum charges, associated with
$\partial_t,\partial_\phi$, are excited in our solution. The
overall additive constant is fixed by the condition that charges
vanish in the regular $\alpha=1, A=0$ vacumm metric. In addition
an overall ${1\over g_6^2\, V_{S^3\times S^1}}={1\over 4\,\pi^3\,
g_6^2}$ normalization factor is included in the definition
(\ref{Q}) to account for the difference between the string and
Einstein metrics and the normalization to one of the volume of the
$S^3\times S^1$ boundary in the case of spheres of unit radii. The
final result can be written as:
\begin{eqnarray}
\mathrm{J}[\ell\,\partial_t]&=& {\hat{l}^4\over 2}\,
\left(1-\alpha^2-\frac{A^2}{\ell^2}\right)\nonumber\\
\mathrm{J}[\ell\, \partial_\phi]&=&2\,\hat{l}^4  \,A\,\alpha
\end{eqnarray}
where $\hat{l}^4={\ell^4\over g_6^2}=Q$ reabsorbs the ${1\over
g_6^2}$ factor in front of the charge definition. We notice that
$b$ does not enter the above expression for the charges.

As promised we now evaluate the central extension of
the Virasoro algebras realized by (\ref{ln}).
Following \cite{bh} we can derive this term by
evaluating the Poisson brackets:
\begin{eqnarray}
\left[\mathbb{J}[\xi],\,\mathbb{J}[\eta]\right]&=&
\delta_\eta\,\mathbb{J}[\xi]
\label{poisson}
\end{eqnarray}
with $\xi=L_n$ and $\eta=L_{-n}$ given by:
\begin{eqnarray}
  L_n&=&\frac{i \,\ell}{2}\, e^{ i\,n\,(\frac{t}{\ell}-\phi)}\,
\left[ \left(1 - {n^2\,\ell^2\over 2\, r^2} \right)
  \,\partial_t -\frac{1}{\ell}\,\left(1 + {n^2\,\ell^2\over 2\, r^2} \right)
  \,(\partial_\phi-\ell\,b \, \partial_{\tilde{\varphi}_2}) -\frac{i\,n\, r}{\ell}\,\partial_r
\right] \nn\\
  \bar{L}_n&=&\frac{i \,\ell}{2}\, e^{ i\,n\,(\frac{t}{\ell}+\phi)}\,
\left[ \left(1 - {n^2\,\ell^2\over 2\, r^2} \right)
  \,\partial_t +\frac{1}{\ell}\,\left(1 + {n^2\,\ell^2\over 2\, r^2} \right)
  \,(\partial_\phi-\ell\,b \, \partial_{\tilde{\varphi}_2}) -\frac{i\,n\, r}{\ell}\,\partial_r
\right]
\label{ln}
\end{eqnarray} The variations of the metric
$\delta_{\xi}\, h_{ab}$
and canonical momentum $\delta_{\xi}\, h_{ab}$ are now defined by
the Lie derivative $\delta_{\xi}={\cal L}_{\xi}$ along
the vector field $\xi$.
  Plugging (\ref{hn2}) in (\ref{dJ}), after
some algebra one is left with the final result \be
\left[\mathbb{J}[\xi],\,\mathbb{J}[\eta]\right]= -{{\rm i} \,
n\over 2}\, \hat{\ell}^4 \, \left(\alpha^2+{A^2\over \ell^2}+{4\,
A\,\alpha\over \ell} -n^2\right) \label{poissonf} \ee One can see
that the result (\ref{poissonf}) can be written as a sum of two
pieces: \bea
 2\, n\, \mathbb{J}[L_0] &=&i\,\ell\, n\,
\left(\mathbb{J}[\partial_t]-\ell\,\mathbb{J}[\partial_\phi]\right)\nn\\
&&=- {{\rm i}\, n\, \hat{\ell}^4\over 2}\,
\left(\alpha^2+{A^2\over \ell^2}+{4\,A\,\alpha\over
\ell}-1)\right) \eea and the central term ${{\rm i}\over
2}\,\hat{\ell}^4\, (n^3-n)$. We conclude that the asymptotic
Killing generators realize two copies of Virasoro algebras with
commutation relations:
\begin{eqnarray}
\left[\,\mathbb{J}[L_n] ,\,\mathbb{J}[L_m] \,\right]&=&
(n-m)\,\mathbb{J}[L_{n+m}] +{{\rm i}\,c\over 12}\, (n^3-n)\,\delta_{n+m,\,0}
\nonumber\\
\left[\,\mathbb{J}[\bar{L}_n] ,\,\mathbb{J}[\bar{L}_m] \,\right]&=&
(n-m)\,\mathbb{J}[\bar{L}_{n+m}] +{{\rm i}\,c\over 12}
\,(n^3-n)\,\delta_{n+m,\,0}\nonumber\\
\left[\,\mathbb{J}[L_n] ,\,\mathbb{J}[\bar{L}_m] \,\right]&=&0
\label{Virasoro}
\end{eqnarray}
with the expect central charge $c=6\,\hat{\ell}^4=6\, Q$.
Notice that the result is independent of the flux parameter $b$.

\section{Concluding remarks}

In this paper we construct solutions of five-dimensional
supergravity which provide a simple setting where the physics of
RG-flows out of two-dimensional $N=(4,0)$ CFTs can be
quantitatively studied. The solutions are constructed by tilting
$AdS_3\times S^3$ geometries and further reducing it to
five-dimensions. The final results are warped geometries of Melvin
type with various NS--NS/RR fluxes.

We identify interesting decoupling limits at the two ends of the
flow. In both cases the  deep inside  region can be accurately
described by an $AdS_3\times S^2$ exact solution of
five-dimensional supergravity. This background corresponds to  the
reduction from the
 $AdS_3\times S^3/Z_N$ type IIB vacuum  on the Hopf fiber of the compact factor.
In the ultraviolet, the two dimensional theory flows to a
non-conformal
 theory with non-trivial five dimensional dilaton,
 better described in terms of the familiar
 six-dimensional $AdS_3\times S^3$ supergravity vacuum
with constant dilaton. The $S^2$ and $S^3$ isometries realize the
global part of the ${\cal N}=(4,0)$ and ${\cal N}=(4,4)$ conformal
field theories at the two ends of the flow. We stress the fact
that the two $AdS_3$ are relatively tilted due to the non-trivial
global identifications that mix the $AdS_3$ and $S^3$ geometries,
making a purely three dimensional analysis along the lines of
\cite{bs} more involved.
   In these more conventional terms the flux
solutions above correspond to the flow out of an $AdS_3$ vacuum of
three-dimensional $SU(2)$ gauged supergravity with ${\cal N}=4$
unbroken supercharges, towards an asymptotic geometry with
non--trivial dilaton.
 It would be very interesting, to apply the techniques
developed in \cite{bs}, to make this correspondence more precise.
 Alternatively one can study the flow out of the $AdS_3\times S^3$
fixed point geometry in the asymptotic region using the tools of
${\cal N}=8$ gauged supergravities \cite{ads3}.

In \cite{ghmn}, the spectrum of D1D5 BPS excitations in various
freely acting orbifolds and orientifolds of type IIB theory was
determined and they were shown to match the AdS/CFT predictions in
terms of chiral harmonics of the corresponding $AdS_3\times S^3$
dual supergravities (see \cite{boer} for earlier results in the
more familiar D1D5 systems for type IIB on $T^4$ or $K3$ and
\cite{dps} for a D1D5 system with $AdS_3\times S^3\times S^3$ dual
supergravity).
 It would be interesting to apply these techniques to provide more
quantitative tests of the correspondences proposed here.

   Another interesting feature of the $AdS_3\times S^2$ decoupled
geometry is that it provides a black string solution in $D=5$
dimensions which can be used as the starting point for the
construction of four-dimensional black holes \footnote{Making
contact for instance with the results of \cite{RR} in which a
systematic geometrical analysis of the microscopic
 description of BPS black holes in $N=8\, D=4 $
supergravity was carried out.}.

   Finally it
would be nice to extend the results attained in the present work
to
 cases involving four-dimensional gauge theories.
We believe that this analysis can provide a deep insight into the
nature of the lifts of locally $AdS_5$ geometries, associated with
${\cal N}=1,2$ gauge theories, to ten dimensions. Another
possibility is the study of the effects of introducing
supersymmetric RR flux--branes on $AdS_5\times S^5$.

 We plan to deal with some of these issues
in the near future.

\vspace*{1cm}

{\large{\bf Acknowledgements}}

\vspace*{0.3cm}
\noindent
We thank G. Bonelli, L. Cornalba J. de Boer  and
H. Samtleben for useful discussions. This work is supported
in part by the European Union RTN contracts
HPRN-CT-2000-00122 and HPRN-CT-2000-00131.

\end{document}